\documentclass{article}
\usepackage{spconf,preamble}
\usepackage{graphicx,hyperref}
\hypersetup{
	colorlinks=true,
	linkcolor=blue,
	citecolor=green!50!black!,
	urlcolor=magenta
}
\usepackage{cite}

\usepackage{xcolor} 
\usepackage{enumitem} 
\usepackage{amsthm} 

\usepackage{tikz}
\usetikzlibrary{positioning,arrows.meta,fit,calc,shapes.arrows}

\usepackage{algorithm} 
\usepackage{algorithmic} 

\title{Learning to Replace MCMC in Split-Gibbs Diffusion\\ Posterior Sampling via Deep Unfolding}
\name{Yi Zhang$^{1}$, Rui Guo$^{1}$, Mengchu Xu$^{1}$, Zhaofeng Liu$^{2}$, Yonina C. Eldar$^{1,2}$ \thanks{This research was supported by the European Research Council (ERC) under the European Union’s Horizon 2020 research and innovation program (grant No. 101000967) and by the Israel Science Foundation (grant No. 536/22). This work is also supported by Manya Igel Centre for Biomedical Engineering and Signal Processing.}
}
\address{$^{1}$Faculty of Mathematics and Computer Science,
Weizmann Institute of Science, Rehovot, Israel\\
$^{2}$Department of Electrical and Computer Engineering,
Northeastern University, Boston, MA, USA}

\begin{document}
%
\maketitle
\begin{abstract}
	Split Gibbs sampling enables diffusion posterior inference for general nonlinear inverse problems by decoupling prior and likelihood computations, allowing a pretrained diffusion prior to be reused across measurement models. However, its likelihood update often relies on iterative MCMC, which can hinder parallelization, require algorithm-specific tuning, and incur substantial computational cost. In this work, we propose a learning-based framework to replace this MCMC step by reformulating both Gibbs updates as Gaussian denoising problems and implementing them through ODE diffusion. The prior step reuses a pretrained denoiser, while the likelihood denoiser exploits known likelihood structure through a lightweight deep-unfolded network. Experiments on nonlinear phase retrieval demonstrate the effectiveness of the proposed method as an alternative to MCMC-based split Gibbs at lower likelihood-update cost.
\end{abstract}

\section{Introduction}
\label{sec:intro}


Diffusion models have emerged as powerful pretrained priors for inverse problems \cite{chung2023diffusion,zhang_improving_2025,daras_survey_2024,zhang2026icassp}. Owing to their ability to generate high-quality samples from complex data distributions, diffusion models facilitate sampling from the posterior $p_{S|Y}(\cdot|y)\propto p_{S}(\cdot) \times p_{Y|S}(y|\cdot)$, where $S$ is the unknown signal and $Y=y$ is the observation, allowing various posterior statistics to be estimated from generated samples.

Split Gibbs sampling \cite{vono2019,coeurdoux2024,xu2024,wu_principled_2024,guo2025,gao2025,zhang_bayesian_2026} provides a flexible framework for diffusion posterior inference in general nonlinear inverse problems, by introducing an auxiliary signal variable \(S'\) to decouple the prior and likelihood factors. Specifically, it constructs the augmented joint distribution
\begin{equation}\label{eq:augment-joint}
	\pi_\tau(s,s') \propto p_S(s)\times p_{Y|S}(y|s')\times \exp\paren{ -\inv{2\tau^2} \norm{s'-s}^2_2 }
\end{equation}
whose $S$-marginal approaches the target posterior as $\tau\to 0$. Here, since the prior factor $p_S(s)$ and likelihood factor $p_{Y|S}(y|s')$ depend on separate variables and interact only through the Gaussian coupling, alternating Gibbs updates of $S$ and $S'$ separate the prior- and likelihood-dependent computations. In diffusion-based implementations, these two updates can be handled independently by a pretrained diffusion model and a conventional MCMC sampler \cite{coeurdoux2024,xu2024,wu_principled_2024,guo2025,gao2025,zhang_bayesian_2026}, respectively, allowing the same diffusion prior to be reused across different inverse problems without retraining.

However, the likelihood-side MCMC introduces several practical limitations. First, the number of iterations required for adequate mixing generally varies across samples, making adaptive computation difficult to parallelize, while a fixed budget may sacrifice either efficiency or sampling quality. Second, MCMC performance can be sensitive to algorithm-specific choices such as stepsizes and proposal scales, introducing additional tuning effort \cite{brooks_handbook_2011}. Beyond these implementation issues, the repeated inner sampling itself can incur substantial computational cost, particularly when likelihood evaluation is expensive or the corresponding conditional distribution is difficult to explore.

In this work, we propose a learning-based framework for replacing iterative likelihood-side MCMC in split-Gibbs diffusion posterior sampling while preserving modular prior-likelihood computation. Our approach progressively reduces the likelihood-side sampling problem to a lightweight unfolded denoiser through three key ingredients:
\begin{enumerate}[label=\arabic*)]
	\item \textbf{Alternating-denoising formulation of split Gibbs.} We cast both the prior and likelihood updates of split Gibbs as Gaussian denoising sampling problems, yielding an exact symmetric formulation equivalent to the original split Gibbs sampler and thus preserving its modular structure and theoretical guarantees.
	\item \textbf{ODE-diffusion implementation of denoising sampling.} We implement both denoising updates through ODE-based diffusion, reducing them to evaluations of prior and likelihood denoisers. The prior denoiser can be directly reused from a pretrained diffusion model, leaving the likelihood denoiser as the only problem-specific learned component.
	\item \textbf{Lightweight likelihood denoising via deep unfolding.} Rather than designing a generic high-dimensional conditional denoiser, we exploit the known likelihood structure to derive a deep-unfolded architecture from a model-based iterative solver, yielding a structured low-parameter network.
\end{enumerate}
We validate the proposed method on nonlinear phase retrieval, where the learned likelihood updates achieve reconstruction quality comparable to their MCMC-based counterparts with a \(4.98\times\) speedup in likelihood-update computation. These results support the proposed method’s potential as an efficient alternative to MCMC-based likelihood updates.

\section{Split Gibbs as an Alternating Denoising Conditional Sampler}


In this section, we first review the standard split-Gibbs posterior sampler and its alternating prior and likelihood updates. We then show that both updates admit a common Gaussian-denoising formulation, yielding a symmetric alternating-denoising sampler. This formulation enables both updates to be implemented within a common ODE-diffusion framework, as described in the next section.

\subsection{Standard Split Gibbs Posterior Sampling}
\label{sec:split-Gibbs}

Given an observation $Y=y$, split Gibbs sampling performs posterior inference by applying Gibbs sampling to the augmented joint distribution \eqref{eq:augment-joint}. Specifically, $S$ and $S'$ are updated alternately by sampling from the conditional distributions $\pi_\tau(s|s')$ and $\pi_\tau(s'|s)$, respectively, yielding the following prior (P-step) and likelihood (L-step) updates \cite{vono2019,coeurdoux2024,xu2024,wu_principled_2024}:
\begin{align*}
	\textbf{P-step:}\; & \text{sample}~S_i \sim  c \times p_S(\cdot)\times \exp\paren{ -\frac{ \norm{\cdot - S'_{i-1} }^2_2 }{2\tau_i^2} },      \\
	\textbf{L-step:}\; & \text{sample}~S'_i \sim c'\times p_{Y|S}(y|\cdot) \times  \exp\paren{ -\frac{ \norm{\cdot - S_{i} }^2_2 }{2\tau_i^2} },
\end{align*}
where $c,c'>0$ are normalizing constants, and the coupling strength $\tau$ in \eqref{eq:augment-joint} is replaced by an iteration-dependent sequence $\paren{\tau_i}_{i=1}^{\infty}$.

The P-step and L-step involve only the prior factor $p_S(\cdot)$ and the likelihood factor $p_{Y|S}(y|\cdot)$, respectively. In practice, the prior distribution often lacks a tractable analytical form. When a pretrained denoising network for $p_S$ is available, however, the P-step can be implemented using diffusion-based sampling \cite{coeurdoux2024,xu2024,wu_principled_2024,guo2025,gao2025,zhang_bayesian_2026}. In contrast, the likelihood $p_{Y|S}$ is typically specified explicitly by the observation model, making conventional MCMC methods applicable to the L-step \cite{xu2024,wu_principled_2024}.

Importantly, under suitable regularity conditions, if the P- and L-steps exactly sample from their respective conditional distributions and the coupling parameter $\tau_i$ decreases sufficiently slowly to zero, the distribution of $S_i$ converges to the target posterior $p_{S|Y}(\cdot|y)$ \cite{vono2019,xu2024}.

\subsection{Reformulating P- and L-Steps as Denoising Samplers}
In standard diffusion-based implementations of split Gibbs, the P- and L-steps are handled asymmetrically. We now show that both steps can be cast as Gaussian denoising sampling problems, yielding a symmetric reformulation of split Gibbs. To this end, we first introduce notation for a Gaussian denoising conditional distribution.

For a probability distribution $\mu$ with density, also denoted by $\mu$, a noise level $\tau>0$, and a noisy observation $z$, we define
\begin{equation}\label{eq:denoise-pdf}
	g_{\tau}[\mu](\cdot|z)
	\triangleq
	p_{X|X+\tau N}(\cdot|z),
\end{equation}
where $X\sim\mu$ and $N\sim\mathcal{N}(\mathbf{0},\mathbf{I})$ is independent of $X$. Applying Bayes' rule to \eqref{eq:denoise-pdf} yields
\begin{equation}\label{eq:denoise-pdf-expansion}
	g_{\tau}[\mu](\cdot|z) \propto \mu(\cdot) \times \exp\paren{- { \norm{\cdot-z}^2_2 }/\paren{2\tau^2} }
\end{equation}
By comparing this expression with the conditional distributions defining the P- and L-steps, we can rewrite both steps as sampling from $g_\tau[\mu](\cdot|z)$ with suitable $\mu,\tau,z$.

\begin{figure*}[!t]
	\centering
	\begin{tikzpicture}[
			block/.style={
					draw,
					thick,
					rectangle,
					align=left,
					inner sep=6pt
				},
			subblock/.style={
					draw,
					thick,
					rectangle,
					align=center,
					inner sep=4pt
				},
			>=Latex
		]


		\node[block, text width=6.5cm] (pstep) {
			\textbf{P-step:}\\
			Sample $\bar X_T\sim\mathcal N(0,\sigma(T)^2I)$\\
			Set $\tau\triangleq\tau_i,\; z\triangleq S'_{i-1},\;
				D_\theta(x,\eta)\triangleq D_\theta^{\mathrm P}(x,\eta)$\\
			Simulate~\eqref{eq:denoise-ODE} from $t=T$ to $0$\\
			$S_i\gets\bar X_0$
		};

		\node[block, text width=8cm, right=0.8cm of pstep] (lstep) {
			\textbf{L-step:}\\
			Sample $\bar X_T\sim\mathcal N(0,\sigma(T)^2I)$\\
			Set $\tau\triangleq\frac{\tau_0\tau_i}
				{\sqrt{\tau_0^2-\tau_i^2}},\;
				z\triangleq\frac{\tau_0^2S_i}
				{\tau_0^2-\tau_i^2},\;
				D_\theta(x,\eta)\triangleq
				D_\theta^{\mathrm L}(x,\eta;y)$\\
			Simulate~\eqref{eq:denoise-ODE} from $t=T$ to $0$\\
			$S'_i\gets\bar X_0$
		};

		\draw[->, thick]
		([xshift=-0.8cm]pstep.west)
		-- node[above, midway, yshift=1pt, xshift=-3pt]
		{$S'_{i-1}$}
		(pstep.west);

		\draw[->, thick]
		(pstep.east)
		-- node[above, midway, yshift=1pt]
		{$S_i$}
		(lstep.west);

		\draw[->, thick]
		(lstep.east)
		-- node[above, midway, yshift=1pt, xshift=-3pt]
		{$S'_i$}
		++(0.8cm,0);


		\node[
			subblock,
			below=1.55cm of pstep.south,
			xshift=2cm
		] (grad) {
			likelihood-gradient step
		};

		\node[
			subblock,
			right=0.45cm of grad
		] (mlp) {
			component-wise MLP
		};

		\node[left=1.2cm of grad] (dotsL) {$\cdots$};
		\node[right=1.2cm of mlp] (dotsR) {$\cdots$};

		\coordinate[left=1.75cm of dotsL] (start);
		\node[right=1.20cm of dotsR] (dout)
		{$D_\theta^{\mathrm L}(x,\eta;y)$};

		\draw[->, thick]
		(start)
		-- node[above, midway] {$s'_0\triangleq x$}
		(dotsL.west);

		\draw[->, thick]
		(dotsL.east)
		-- node[above, midway] {$s'_k$}
		(grad.west);

		\draw[->, thick]
		(grad.east)
		-- (mlp.west);

		\draw[->, thick]
		(mlp.east)
		-- node[above, midway] {$s'_{k+1}$}
		(dotsR.west);

		\draw[->, thick]
		(dotsR.east)
		-- node[above, midway] {$s'_K$}
		(dout.west);

		\coordinate (ytop) at ([yshift=0.62cm]grad.north);
		\draw[->, thick]
		(ytop)
		-- node[left=1pt, midway] (ylabel) {$y$}
		(grad.north);

		\coordinate (xetatop) at ([yshift=0.62cm]mlp.north);
		\draw[->, thick]
		(xetatop)
		-- node[left=1pt, midway] (xetalabel) {$(x,\eta)$}
		(mlp.north);

		\node[
			draw,
			thick,
			dashed,
			rectangle,
			fit=(grad)(mlp)(ytop)(xetatop)(ylabel)(xetalabel),
			inner xsep=0.18cm,
			inner ysep=0.16cm
		] (unfolded) {};

		\node[
			font=\bfseries,
			fill=white,
			inner xsep=4pt,
			inner ysep=0pt
		] at (unfolded.north) {
			$k$-th unfolded layer
		};

		\coordinate (lattach) at (dout.north |- lstep.south);

		\node[
			single arrow,
			single arrow head extend=2pt,
			draw=blue!70!black,
			fill=blue!55,
			minimum width=0.28cm,
			minimum height=0.78cm,
			inner sep=0pt,
			rotate=90
		] at ($(dout.north)!0.52!(lattach)$) {};

	\end{tikzpicture}

	\vspace{-0.5em}
	\caption{Deep-unfolded ODE-diffusion implementation of the alternating-denoising split Gibbs sampler.}
	\label{fig:framework}
\end{figure*}
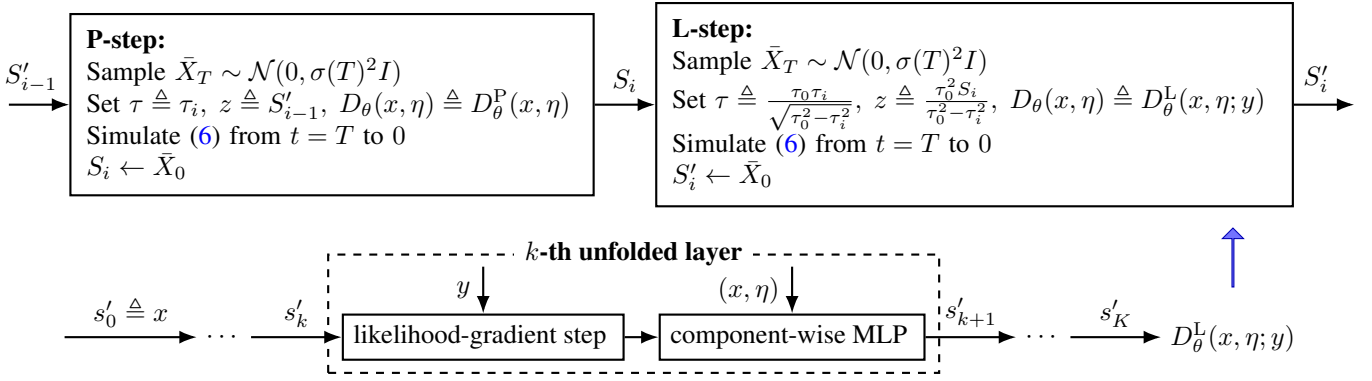

\textbf{Reformulating the P-step}. From \eqref{eq:denoise-pdf-expansion}, we have
\begin{equation*}
	g_{\tau_i}[p_S](\cdot|S'_{i-1}) \propto p_S(\cdot)\times \exp\paren{ -{ \norm{\cdot - S'_{i-1} }^2_2 }/\paren{2\tau_i^2} }.
\end{equation*}
Hence, the P-step is exactly a denoising sampling step.

\textbf{Reformulating the L-step.}
Unlike $p_S(\cdot)$, the likelihood factor $p_{Y|S}(y|\cdot)$ is generally not a probability density and cannot be directly used as $\mu$ in \eqref{eq:denoise-pdf-expansion}. To handle $p_{Y|S}(y|\cdot)$ through the same Gaussian-denoising construction as $p_S$ in the P-step, we introduce a surrogate observation model that replaces the true prior with a fictitious Gaussian prior:
\begin{equation}\label{eq:surrogate-model}
	S' \xrightarrow{p_{Y|S}} Y',~\text{where}~S'\sim \Normal{\mathbf{0}}{\tau_0^2\Imat},
\end{equation}
where $\tau_0$ is a user-specified hyperparameter satisfying $\tau_0>\tau_i$ for all $i$. Using the surrogate posterior $p_{S'|Y'=y}$ as $\mu$ in $g_\tau[\mu](\cdot|z)$, the L-step can then be written as a Gaussian-denoising step:
\begin{align*}
	 & g_{\frac{\tau_0\tau_i}{\sqrt{\tau_0^2-\tau_i^2}}}
		[p_{S'|Y'=y}]
	\cparen{\cdot}{
		\frac{\tau_0^2}{\tau_0^2-\tau_i^2}S_i
	}
	\\
	 & \qquad\propto
	p_{Y|S}(y|\cdot)
	\times
	\exp\paren{
		-\frac{\norm{\cdot-S_i}_2^2}{2\tau_i^2}
	}.
\end{align*}
Combining the two reformulations above yields the \textbf{alternating-denoising form of the split-Gibbs iteration}:
\begin{equation}\label{eq:denoising-split-Gibbs}
	\begin{aligned}
		\textbf{P-step:}\; & \text{sample}~S_i \sim g_{\tau_i}[p_S](\cdot|S'_{i-1}),                                                                                               \\
		\textbf{L-step:}\; & \text{sample}~S'_i\sim g_{\frac{\tau_0\tau_i}{\sqrt{\tau_0^2 - \tau_i^2}}}[p_{S'|Y'=y}]\cparen{\cdot}{ \frac{\tau_0^2 S_i}{ \tau_0^2 - \tau_i^2 }  },
	\end{aligned}
\end{equation}
which preserves the modular prior-likelihood structure and theoretical guarantees \cite{xu2024} of split Gibbs and enables a common denoising implementation for both updates.

\section{Implementation of the Denoising Conditional Sampler}

We next implement the Gaussian denoising sampling steps in \eqref{eq:denoising-split-Gibbs} in two stages: ODE-based diffusion reduces each sampling step to denoiser evaluations, after which deep unfolding exploits the likelihood structure to derive a principled, lightweight architecture for the likelihood denoiser.

\subsection{ODE-Based Diffusion for Denoising Sampling}

For a general distribution $\mu$, Gaussian denoising sampling can be implemented through an ODE-based diffusion process using a denoiser for $\mu$. Specifically, suppose that a denoising network $D^\mu_\theta(x,\eta)$ satisfies
\begin{equation*}
	D^\mu_\theta(x,\eta)
	\approx
	\Ebb_{X\sim\mu}\cbrac{X}{X+\eta N=x},\;\text{where}~N\sim\Normal{\zerov}{\Imat}.
\end{equation*}
Then, a sample from $g_\tau[\mu](\cdot|z)$ can be generated by simulating the following reverse-time ODE $(\bar{X}_t)_{t\geq 0}$:\footnote{This ODE-based diffusion process was previously derived in \cite{xu2024} for the P-step; our reformulation here extends its applicability to the L-step, enabling a unified diffusion implementation of both steps.}
\begin{equation}\label{eq:denoise-ODE}
	\frac{d\bar{X}_t}{dt}  = \frac{\dot{\sigma}(t)}{\sigma(t)} \paren{ \bar{X}_t - D^\mu_\theta\paren{ \frac{ \tau^2 \bar{X}_t + \sigma(t)^2 z }{ \tau^2 + \sigma(t)^2  } ,  \frac{\tau \sigma(t)}{ \sqrt{\tau^2 + \sigma(t)^2} }  } },
\end{equation}
where $\sigma(\cdot)>0$ is an increasing noise-level function specified by the user, and $\dot{\sigma}(\cdot)$ is its derivative. If the diffusion terminal time $T>0$ is chosen such that $\sigma(T)$ is much larger than the standard deviation of $\mu$, initializing $\bar{X}_T\sim\Normal{\zerov}{\sigma(T)^2\Imat}$ and simulating \eqref{eq:denoise-ODE} backward from $T$ to $0$ yields $\bar{X}_0$ as an approximate sample from $g_\tau[\mu](\cdot|z)$.

Applying this general framework to \eqref{eq:denoising-split-Gibbs} requires two denoisers corresponding to the P- and L-steps. For the P-step, $\mu=p_S$, and hence requires
\begin{equation*}
	D^{\mathrm P}_\theta(x,\eta)
	\approx
	\CondExpt{S}{S+\eta N=x},
\end{equation*}
which can be implemented by a pretrained diffusion denoiser \cite{song_score-based_2021,ho_denoising_2020,karras_elucidating_2022}. For the L-step, $\mu=p_{S'|Y'=y}$ depends on the observation $y$, and therefore requires the conditional denoiser
\begin{equation*}
	D^{\mathrm L}_\theta(x,\eta;y)
	\approx
	\CondExpt{S'}{S'+\eta N=x,Y'=y}.
\end{equation*}
Thus, ODE diffusion reduces the implementation of \eqref{eq:denoising-split-Gibbs} to evaluations of two denoisers: the prior denoiser can be reused from a pretrained diffusion model, whereas the likelihood denoiser must be designed for the given likelihood. The resulting ODE-diffusion framework is shown in Fig.~\ref{fig:framework}; we next construct the latter denoiser via deep unfolding.

\subsection{Likelihood Denoiser Design via Deep Unfolding}

\begin{figure*}
	\centering
	\includegraphics[width=.95\textwidth]{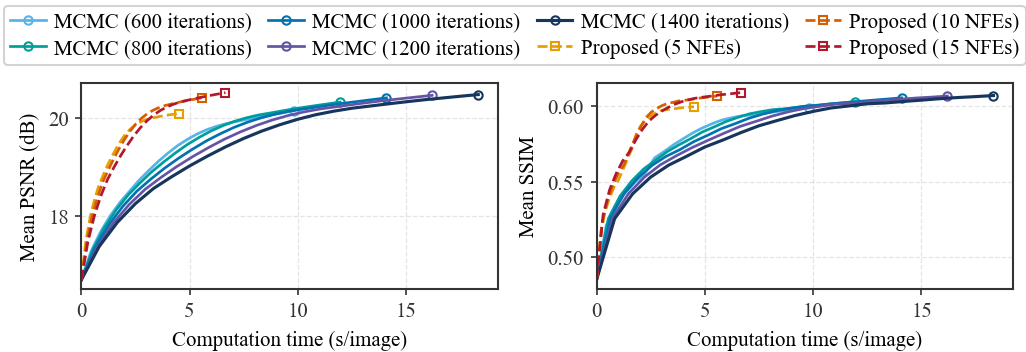}
	\vspace{-0.3cm}
	\caption{Reconstruction quality versus computation time for different implementations of the split Gibbs sampler. Time includes both prior and likelihood updates; markers indicate the endpoints after 22 outer iterations.}\label{fig:quality-vs-time}
\end{figure*}
\begin{table*}[t]
	\centering
	\vspace{-0.4cm}
	\caption{Average likelihood-update time per image per outer iteration and reconstruction quality after 22 outer iterations.}\label{tab:main-results}
	\label{tab:warm500-runtime-quality}
	\setlength{\tabcolsep}{4pt}
	\renewcommand{\arraystretch}{1.12}
	\begin{tabular}{@{}l*{5}{r}|*{3}{r}@{}}
		\hline
		                     & \multicolumn{5}{c|}{MCMC} & \multicolumn{3}{c}{Proposed}                                                                     \\
		\cline{2-6}\cline{7-9}
		Likelihood update    & 600 iter.                 & 800 iter.                    & 1000 iter. & 1200 iter. & 1400 iter. & 5 NFEs & 10 NFEs & 15 NFEs \\
		\hline
		Likelihood time (ms) & 289.964                   & 386.634                      & 483.589    & 579.937    & 676.557    & 48.624 & 97.061  & 145.566 \\
		PSNR (dB) $\uparrow$ & 20.147                    & 20.325                       & 20.412     & 20.463     & 20.479     & 20.092 & 20.412  & 20.518  \\
		SSIM $\uparrow$      & 0.5990                    & 0.6032                       & 0.6056     & 0.6068     & 0.6072     & 0.5997 & 0.6068  & 0.6091  \\
		\hline
	\end{tabular}
\end{table*}

The likelihood denoiser $D^{\mathrm L}_\theta$ maps the high-dimensional inputs $(x,\eta,y)$ to a denoised estimate, for which a generic black-box network offers no natural choice of hidden representation and can lead to a large parameterization. Since the joint distribution of $(S',Y',S'+\eta N)$ is fully specified by the surrogate model \eqref{eq:surrogate-model}, we instead exploit this known structure to construct a lightweight model-informed architecture via deep unfolding \cite{monga_algorithm_2021,shlezinger_deep_2025}.

Specifically, we derive an unfolded proximal-gradient architecture \cite{combettes2005,wei2022} from the MAP estimation of $S'$. The MAP formulation is used only to motivate the network architecture; the subsequent MSE training targets the desired conditional-mean denoiser. Given $S'+\eta N=x$ and $Y'=y$, the MAP estimate solves
\begin{equation*}
	\min_{s'}-\log p_{Y|S}(y|s')
	-\log p_{S'|S'+\eta N}(s'|x).
\end{equation*}
Let us define two auxiliary functions $f_y(s')\triangleq-\log p_{Y|S}(y|s')$ and $g^\eta_x(s')\triangleq-\log p_{S'|S'+\eta N}(s'|x)$, then the corresponding proximal-gradient iteration for solving the MAP problem is
\begin{equation}
	s'_{k+1}
	=
	\operatorname{Prox}_{\gamma_k g^\eta_x}
	\paren{s'_k-\gamma_k\nabla f_y(s'_k)}.
\end{equation}
We unfold $K$ iterations into $K$ network layers and make the stepsizes and measurement-model-induced parameters learnable. Since $S'$ and $N$ have independent coordinates, $g^\eta_x$ is coordinate-wise separable; accordingly, we replace its proximal operator by a small MLP applied independently to each coordinate, yielding a low-complexity model-informed network. The network output is defined as $D^{\mathrm L}_\theta(x,\eta;y)\triangleq s'_K$.

Finally, since $(S',Y',N)$ can be sampled entirely from the surrogate model \eqref{eq:surrogate-model}, arbitrarily many synthetic training samples can be generated. We train $D^{\mathrm L}_\theta$ to predict $S'$ by minimizing the MSE
over the considered noise levels $\eta$
\cite{song_score-based_2021,ho_denoising_2020,karras_elucidating_2022}.






\section{Experiments}

We compare learned and MCMC-based likelihood updates in split Gibbs sampling for phase retrieval on FFHQ $256\times256$ images, following the setup, warm-start procedure, and open-source MCMC baseline of \cite{xu2024}. After warm starting, we perform 22 outer iterations for each implementation. Our learned likelihood denoiser comprises six unfolded proximal-gradient layers, each with separately learned forward and backward masks restricted to the physical support of the true mask used in the measurement, and a four-layer MLP for the proximal step. We vary the number of MCMC iterations or network function evaluations (NFEs) of $D^{\text{L}}_\theta$ per likelihood update.

Fig.~\ref{fig:quality-vs-time} shows reconstruction quality versus computation time, and Table~\ref{tab:main-results} reports likelihood-update time and final quality, averaged over 160 images. Prior updates cost approximately 156 ms/image per outer iteration across configurations and are omitted from the table. As shown in Fig.~\ref{fig:quality-vs-time}, the proposed method reaches high PSNR and SSIM in less time. At 22 outer iterations, 15 NFEs yield slightly higher PSNR and SSIM than those of 1400 MCMC iterations. Table~\ref{tab:main-results} shows that the learned likelihood update with 10 NFEs achieves comparable PSNR and SSIM to those obtained with 1000 MCMC iterations, while accelerating the likelihood update by a factor of 4.98. These results demonstrate the effectiveness of replacing MCMC-based likelihood updates with the proposed method and its potential to reduce likelihood-update computation for complex likelihood models.



\section{Conclusion}

We proposed a learning-based framework to replace likelihood-side MCMC in split-Gibbs diffusion posterior sampling. A Gaussian-denoising reformulation enables both Gibbs updates to be implemented through ODE diffusion, with the likelihood denoiser realized by a lightweight model-informed deep-unfolded network. Experiments on nonlinear phase retrieval demonstrate the effectiveness of the proposed method as an alternative to MCMC-based split Gibbs and its potential to reduce likelihood-update cost.

\vfill\pagebreak

\bibliographystyle{IEEEtran}
\bibliography{refs}

\end{document}